\documentclass[preprint,showpacs,preprintnumbers,amsmath,amssymb]{revtex4}
\begin{document}
\topskip 20mm
\title{Thermodynamics of protein folding: a random matrix formulation}
\author{Pragya Shukla}
\affiliation{Department of Physics,
Indian Institute of Technology, Kharagpur, India.}
\date{\today}

\begin{abstract}


The process of protein folding from an unfolded state to a  biologically 
active, folded conformation is governed by many parameters e.g the sequence 
of amino acids, intermolecular interactions, the solvent, temperature and 
chaperon molecules. Our study, based on random matrix modeling of the 
interactions, shows however that the  evolution of the statistical measures 
e.g Gibbs free energy, heat capacity, entropy  
is single parametric. The information can explain the selection of specific folding 
pathways from an infinite number of possible ways as well as other folding characteristics 
observed in computer simulation studies.

\end{abstract}
                                             


\maketitle

\section {Introduction}


The expression of a gene in a DNA leads to formation of amino acids sequences that are the 
basic building blocks of proteins. The message contained in a DNA then manifest through a 
specific structure of protein which in turn determines its functionality. In fact, the protein  
after its birth, acts as a feedback and leads to creation of new copies of the parent DNA.

The structure of a protein is determined purely by the amino acid sequences and 
its function depends on the ability of the protein to fold rapidly to its native 
structure \cite{rev, ri}. Based on numerous simulation studies of protein sequences 
(for example, see \cite{rev, ri, num1, num2, lee, sho, hao1, shak2, shak3, cam, leo}), 
the folding process is believed  to reveal two main characteristics: 
(1) a single thermodynamically stable, minimum free energy state,
(2) a very short time-scale for folding e.g. milliseconds to seconds.
In past, there have been several analytical attempts to explain 
these observations (see for example 
\cite{ hao1, hoang, seno, ana, bry1, shak1}). 
However a thorough understanding of the rapid 
and selective approach of a  sequence to fold to a pre-determined 
configuration, despite availability of an infinite number 
of possibilities, is still missing (referred as protein folding problem). 
The three main components of the missing information are:
(i) an understanding of the inter-atomic forces which lead to native 
state from an unfolded state, 
(ii) prediction of native structure from its amino acid sequences 
(usually requires a prior knowledge of inter-atomic forces),
(iii) the origin/ reason of fast folding speed.
We seek the information by a new analytical method based 
on the random matrix modeling \cite{meta} of the interactions within 
protein as well as with its environment, and attempt to justify the findings of 
the simulation studies.

The interactions among various units of a biological system are often complicated and can not be determined exactly. The  complexity of the  interactions manifests  itself through sample to sample fluctuations of the properties. Such fluctuations (different from thermodynamic or statistical ones)  have been observed in a wide range of complex systems and a useful information can be extracted only from the statistical analysis of properties \cite{meta}. For example, the microscopic energy states  of complex systems like proteins are not well-defined and can at best be described by a statistical distribution. Previous analytical studies attempted to circumvent this difficulty 
by  averaging over the ensemble of protein sequences and, therefore,  could 
not provide information about the role of a specific sequence \cite{hao1} on the folding.  Our approach however is based on the averaging over the 
ensemble of interactions of a given sequence and does not suffer from  
this drawback. We analyze the interaction matrix i.e 
the matrix with its entries as the pairwise interactions between residues as 
well as their side-chains of a given sequence.   The deterministic inaccuracy associated with the interactions results in their distribution (spread about 
some average value), with nature and degree of randomness 
governed by the local environment \cite{ps-valla}.  The interaction matrix then turns out to be a random matrix i.e a matrix with some/all random entries.  The physical 
properties of such a matrix can be analyzed through 
their ensemble. 

	The concept of randomization of local interactions is essentially 
same in spirit as the idea of randomization of microscopic energy states,  
used in well-known random energy model  of disordered systems \cite{derrida}. The details and the information contained in random matrix model however is significantly different from that of the random energy model.  The latter directly assumes a microscopic energy state to be Gaussian distributed, with all the system-specific information contained in its mean and variance. But the explicit  dependence of mean and variance on the system parameters, e.g. pairwise interaction strengths, is not known 
which reduces the applicability of  the model in probing the folding problem.  
Further, the assumption of randomness  in this case requires presence of disorder. In contrast, the random matrix model, based 
on the inaccuracy led randomization of  local interactions, depends on many parameters, each being a  measure of local interaction-accuracy which in turn is sensitive to the 
system conditions. This  leads to a multi-parametric  distribution of  the microscopic energy states which allows one to explore the effect of local variations on the 
sequence.  Although our analysis finally leads to a single parametric formulation of the energy states,  the parameter is a well-defined functional of the system conditions.    
This makes the model  more appropriate for the analysis of various folding stages (each described by a set of the system parameters).

A protein in aqueous solution is in equilibrium between its native (folded) and 
denatured (unfolded) conformations. The thermodynamic stability of the native state is based on the magnitude of 
the Gibbs free energy $G$ of the system relative to unfolded state. A negative $\Delta G=G_f-G_u$ 
(subscripts $f, u$ implying folded and unfolded state)
indicates the native state is more stable than the denatured one. Many factors are responsible 
for the folding and stability of native proteins, e.g. hydrophobic interactions, hydrogen bonding, 
Van Der Waals forces and electrostatic interactions, conformational entropy, and the physical environment 
(pH, buffer, ionic strength, excipients etc.) \cite{fr, priv, chap}.
The factors  stabilizing the folded state are present in the unfolded state too and  help in its 
stability. The folded state is however marginally more stable than the
unfolded state due to various compensating factors enhancing its stability.  
Further the functionality and folding speed (to native conformation), instead of the thermodynamic stability,  
seem to be the main criteria for the selection of a natural protein conformation. 
Both of these characteristics require some degree of flexibility which in turn 
affects  the free energy constraints on unfolding and refolding.
These insights in  the folding process are
mostly based on computer simulation studies and  it is desirable to seek an
analytical understanding which could then help e.g in designing proteins.  
 This motivates us to consider the partition function of a protein sequence which 
can be used to  determine the stability measure i.e Gibbs free energy of the 
sequence in a specific conformation as well as the heat capacity and entropy 
of unfolding.


During past few decades, the attempts to explain  folding and organization 
of proteins from the unfolded or random coil state to the native folded state 
have put forward  many ideas. It is now believed that the polarity 
of proteins and their hydrophobic interaction with the solvent dominate 
the folding process. The hydrophilic nature of  polar amino acids in 
aqueous solution attracts polar water molecules while non-polar amino 
acids tend to be hydrophobic and prefer binding with each other. 
These tendencies along with other factors confine the space of
available conformations and the folding occurs only through specific pathways. 
It appears to proceed from a restricted conformation
ensemble by condensation and secondary structure formation through an 
even smaller ensemble of "molten globules" to a well-defined, three 
dimensional  single structure \cite{rev, ri}. The final stages of folding 
are also believed to depend on the specific sequence of amino acids, whereas 
earlier stages should be mostly insensitive to the sequence-details. 
Further, molecules of the same protein can follow different pathways to 
reach native state however the thermodynamic stability criteria (requiring 
decrease of free energy) restricts the allowed pathways. To understand these 
pathways,  it is necessary to know the effect of varying residue-residue interactions as well as protein-solvent interactions on the thermodynamic properties. For this 
purpose, we  analyze the energy distribution of a protein sequence under varying 
system conditions which leads to system-dependent formulation of thermodynamic measures.

The paper is organized as follows. The section 2 describes the energy formulation 
for a microscopic state corresponding to a specific conformation. The random matrix 
model of the interactions, based on maximum entropy principle \cite{balian}, is discussed in section 3 which is used in section 4 to obtain the energy 
landscape i.e the distribution of a microscopic state as a function 
of system and environmental conditions. This information 
is applied in section 5 to derive the partition function and Gibbs free energy for folding.  
The  heat capacity for denaturation and thermodynamic entropy are discussed in section 6. The section 7 
contains concluding remarks.

\section{Microscopic energy states of a protein sequence}

A physicist's approach to folding problem is based on applying statistical 
energy functions to explore a large set of alternative structures of a target protein, 
with native state given by the lowest energy structure. An accurate description 
of the Gibb's free energy function needs to take into account the many body interactions among residues, 
(Hydrogen bonds, ion pairs, van Der Waals interactions, hydrophobic interactions) as well as 
effect of the solvent. Fortunately, however, a simplified version of energy function based 
on pairwise contact approximation has turned out to be quite a good description in many 
folding simulation studies \cite{ven, ven1}. Within this approximation,  the energy of a 
particular conformation of a protein sequence of $N$ residues can be 
expressed in terms of a $N \times N$ contact map matrix $C$ whose matrix elements 
represent the pairwise contact potential: 
Consider a sequence $A=(A_1, A_2, A_3....A_N)$,  with $A_k$ as the amino acid at the $k^{th}$ position 
in the chain, folds into a structures whose contact map is $C$. The energy of the conformation 
can be given as 
\begin{eqnarray}
E(C,A,U)= \sum_{kl} C_{kl} U_{kl} (A_k, A_l)  = {\rm Tr}\left[ C . U \right]
\label{e1}
\end{eqnarray}
with $A_k$ as the amino acid at the $k^{th}$ position in the chain. 
Here $U$ is a $N \times N$ symmetric matrix with its elements $U_{kl}=U(A_k,A_l)$ as the 
interaction between residues $A_k$ and $A_l$ (present at position $k$ and $l$ of the sequence), 
where $A_k, A_l$ belong  to a set of the twenty types of amino acids. 

The contact matrix $C$ contains information about the connectedness of  the sequence. 
Based on the connectivity between two residues, the elements of the contact matrix  are 
usually allowed to take binary values: 

\begin{eqnarray}
C_{ij}  &=&   
 1   \qquad {\rm if \; residues \; k \; and \; l \; are \; connected} \nonumber  \\
          &=&    0 \qquad   {\rm otherwise}
\label{e3}
\end{eqnarray}    
The criteria  for connectedness  is usually considered  to be the distance of 
the heavy atoms in the two residues: two residues are assumed to be in contact 
if any two heavy atoms belonging to them are closer than a threshold distance 
( $~ 1-10$ Angstroms).

The effective energy can be rewritten as  

\begin{eqnarray}
E(C,A,U)= {\rm Tr}\left[ H \right]
\label{e4}
\end{eqnarray}
where matrix $H$ is the product of contact matrix $C$ and interaction matrix $U$:
\begin{eqnarray}
H_{kl} = \sum_{j=1}^N  C_{kj} \; U_{jl} 
\label{e5}
\end{eqnarray}

Eq.(\ref{e4}) can be applied to derive $P(E,C,u)$, the distribution of energy 
state $E$ for a specific $C$ matrix, or, the energy landscape for each state of protein 
e.g neutral, charged, folded, intermediate or unfolded
(the energy of a protein being a function of the topological arrangement of the atoms) \cite{fr}. 
An energy landscape depicts energy as a function of the conformation for a given 
state of protein. The stable conformation corresponds to the  global minimum of 
the landscape, with its smooth, well-correlated structure indicating the 
stability of the protein \cite{fr}.

The energy function given in eq.(\ref{e4} ) is one of the most studied
forms  in computer simulation studies of protein folding. Although this 
function is good enough for threading set simulations,  it is
believed to be not accurate enough to allow off-lattice folding
simulations \cite{ven2, ven3}.
This motivated considerations of new energy functions e.g. THOM2
which captures the environment of each residue by assigning a
potential energy $U(A_l, S_{l \alpha})$ for each contact $S_{l \alpha}$ to a
residue $A_l$ \cite{thom}.
 The total energy of a protein in this case is a sum of the site contributions:
\begin{eqnarray}
E(A,U)= \sum_{l=1}^N  \sum_{l_\alpha=1}^{m_l}  U_{l_\alpha} (A_l, A_{l_\alpha})
\label{e6+0}
\end{eqnarray}
where $l=1 \rightarrow N $ with $N$ as the total number of residue sites in the
sequence, $A_{l_\alpha}$ as the $\alpha^{\rm th}$ contact to the residue at site $l$, 
with $l_m$ as the total number of contacts to the site l.

        The interactions between the side chains of various residues is very crucial to
achieve the 3-dimensional structure of unique folded conformation. 
Such interactions are not taken into account in eq.(\ref{e6+0}). This motivates 
us to consider a generalization of eq.(\ref{e6+0}). 
Let  $U_{l_\alpha, k_\alpha} (A_{l_\alpha}, A_{k_\alpha})$ 
be the interaction strength between side chains $A_{l_\alpha}$  and 
$A_{k_\alpha}$, the total energy of pairwise interactions is then    

\begin{eqnarray}
E(A,U)= \sum_{k, l} \sum_{\alpha=1}^{m+1}  
U_{k_\alpha l_\alpha} (A_{k_\alpha}, A_{l_\alpha})
\label{e6+1}
\end{eqnarray}
Note, here the interaction between the residues is included in the sum 
by treating each residue as a side chain too.  
Due to side chain interactions, the size  of the $U$-matrix is now increased: 
$N_u = \prod_{l=1}^N (l_m + 1)$. 
The  missing/ weak connections among the side-chains of different residues, 
and mutually dependent pairwise interactions within a single side chain,  
may lead to an  {\it effectively} sparse form of $U$ matrix with many correlated 
elements.

To proceed further, we need the information about the interactions among  
residues in the sequence as well as with solvent. In protein simulation studies, 
the information is usually taken from protein data bank. However, as discussed in 
the next section, 
the PDB information is only approximately accurate and can be improved  by taking the error 
into account i.e by considering the distribution of interaction strengths. The latter  
is then used to determine the distribution $P(E)$ and the partition function.     

\section{Distribution of interaction strengths: a random matrix model}

Consider the interaction matrix $U$ of a protein sequence with $N$ residues 
with its elements $U_{kl}$ describing  the pairwise interaction between residues 
for a given set of system conditions. 
For notational simplification, henceforth, we denote 
$U_{kl}$ by $U_{\mu}$ with $\mu \equiv \{kl \}$ as a single index (unless 
details required) which can 
take value from $1 \rightarrow M$. Here $M$ is the total number 
of the distinct matrix elements: $M = N(N+1)/2$.

The presence of environment  adds to 
the degree of complexity of the interactions in the chain. This renders an 
exact determination of $U_{\mu}$ technically difficult and they can be 
determined only within a certain degree of accuracy which, being sensitive 
to local system conditions, varies from element to element. 
Each $U_{\mu}$ can  then be best described by a 
distribution with parameters sensitive to system conditions (see \cite{ps-valla}).

Based on extent of available information about system conditions,
the distribution of each $U_{\mu}$ can be obtained by invoking 
maximum entropy hypothesis \cite{balian}: 
{\it in absence of any further information, the simplest 
and least biased hypotheses is that the system is described by the
distribution $\rho(U)$ that maximizes Shannon's information entropy $S$} where

\begin{eqnarray}
S[\rho(U)] = - \int \rho(U)\; {\rm ln} \rho(U)\; {\rm d}\Gamma
\label{c1}
\end{eqnarray}
with $\Gamma(U)$ as the invariant measure in the $U$-space.
For example, consider the system subjected to following constraints:  
(i) the probability density $\rho(U)$ is conserved  (normalized to unity),
(ii) each $U_{\mu}$ is described by an independent, random distribution 
with its higher order ($> 2$) moments negligible, 
(iii) the mean $<U_{\mu}>= u_{\mu}$ and $2^{nd}$ moment $<U_{\mu}^2>=v^2_{\mu}+u^2_{\mu} $ 
are given by the system conditions.  The maximization of Shannon entropy 
under these constraints leads to a Gaussian distribution of $U_{\mu}$:

\begin{eqnarray}
\rho(U) = \prod_{\mu=1}^M {1 \over \sqrt{2 \pi v^2_{\mu}}} 
{\rm e}^{-{(U_{\mu}-u_{\mu})^2 \over 2 v_{\mu}^2}}
\label{e2}
\end{eqnarray}   
where $u_{\mu}$, the ensemble averaged value of interaction,   
could be taken e.g. from a protein data bank.  Note here assumed randomness of an 
interaction is different from considering a "random" sequence. The 
components of a sequence may be well-defined but  their interactions may 
not be. 

The consideration of more realistic conditions e.g. many body interactions 
would introduce  non-zero correlations among $U_{\mu}$s: 
\begin{eqnarray}
\rho(U, v) =  {\tilde C }\prod_{\mu_1, \mu_2} 
{\rm exp}\left[- v_{\mu_1, \mu_2 } (U_{\mu_1}-u_{\mu_1}) \; ( U_{\mu_2} - u_{\mu_2}) \right]
\label{e65}
\end{eqnarray}
with $v_{\mu_1, \mu_2}$ as the measures of correlations between 
$U_{\mu_2}$ and $ U_{\mu_1}$. However in present study we confine 
our analysis to the independent  case.


The Gaussian distributed $U_{\mu}$ (eq.(\ref{e2})) leads to a Gaussian 
ensemble of  $H$-matrix (from eq.(\ref{e5})): 
\begin{eqnarray}
\rho_H(H, C, u) =  \prod_{\mu=1}^M {1 \over \sqrt{2 \pi  \nu^2_{\mu}}} 
{\rm e}^{-{(H_{\mu}-b_{\mu})^2 \over 2 \nu_{\mu}^2}}
\label{e6}
\end{eqnarray} 
with 
\begin{eqnarray}
b_{\mu} & \equiv & \langle H_{\mu} \rangle = \sum_{j} C_{kj} \; u_{jl}, \nonumber \\   
\nu_{\mu} & \equiv & \langle H_{\mu}^2 \rangle- \langle H_{\mu} \rangle^2 =  
\sum_j C_{kj}^2 \; (u_{jl}^2 + v_{jl}^2)  - b_{\mu}^2 
\label{e6-}
\end{eqnarray} 
As clear, $\rho_H$ contains sequences with different interaction energies for a given 
contact map as well as sequences with different contact maps for a given interaction 
matrix $U$.

The energy function in eq.(\ref{e4}) being widely used in simulation studies, 
it is relevant to consider the energy distribution of a sequence modeled by  the ensemble $\rho_H$:  
\begin{eqnarray}
P(E, C, u) = \int \delta(E - {\rm Tr}\left[ H \right] ) \; \rho_H(H) \; {\rm d}H
\label{e6+}
\end{eqnarray}    
$P(E, C, u)$ contains information about the energy landscape: the existence of 
a clear global minimum of $P(E, C, u)$ in $C$-space for a fixed $u$ (i.e a given 
protein sequence) indicates its foldability, 
with the neighborhood containing information about the low-energy 
alternative conformations.  Note the above formulation also 
allows the possibility to consider a more generalized form of contact matrix.

Eq.(\ref{e6+1}) being closer to realistic proteins, our main interest is to
find $P(E)$ for this case: 
\begin{eqnarray}
P(E,u, v) = \int \delta(E - \sum_{\mu} U_{\mu}) 
\; \rho(U, u, v) \; {\rm d}U
\label{e6+2}
\end{eqnarray}
with $\mu \equiv  \{ k_\alpha, l_\alpha \}$, 
$\sum_{\mu} U_{\mu} \equiv \sum_{k,l, \alpha} U_{k_\alpha,  l_\alpha})$  
 and $\rho(U,v,u)$ as the density of the ensemble of $U$-matrices, each of size 
$N_u$.  Assuming the matrix elements correlations negligible, it can again 
be described by eq.(\ref{e2}) with now $M=N_u (N_u+1) /2$.

Eq.(\ref{e6+2}) can model various protein states e.g folded or unfolded. For example, the interactions between side chains in an unfolded state is much 
weaker in comparison to a folded state. The unfolded state can be described 
by eq.(\ref{e2}) by taking
$u_{\mu} \rightarrow 0$, $v_{\mu} \rightarrow 0$ if $\mu \equiv \{ k_\alpha, l_\alpha  \}$ is such that  
$k \not=l$ and if $A_{k_\alpha}$ and $A_{l_\alpha}$ correspond to the side chains.
For native state, a well-defined three-dimensional structure, a large 
number of $u_{\mu}$s would be non-zero with corresponding $v_{\mu}$ very 
small. Similarly the intermediate folding states would correspond to varying  $(u_{\mu}, v_{\mu} )$-strengths,  based on the sequence and its environment. The transition from unfolded to folded state
can therefore be studied by a variation of these  parameters.

\section{Evolution of P(E) during folding process}

Let us first consider the $P(E)$ given by eq.(\ref{e6+2}). 

As the folding proceeds, the interaction strengths  of residues with each other as well as with local environment change and the residues in the sequence rearrange themselves (dictated by their chemical nature and affinities). 
The folding therefore corresponds to dynamics of the elements $U_{\mu}$ and an  
evolution of $\rho(U)$ in the $U$-matrix space.

The deterministic accuracy of each $U_{\mu}$ also fluctuates rapidly as the folding 
evolves, with different "time-scale" of fluctuations for each matrix element.  
This corresponds to  a change of distribution parameters of the ensemble of the 
interaction strengths of a given sequence. The folding process can then be considered   
as an evolution of the ensemble in  the parametric space. Both describing the 
same process, the parametric space dynamics of $\rho(U,u,v)$ is therefore 
expected to mirror itself in its $U$-space dynamics. This is indeed the case as 
can be seen by a partial differentiation of  eq.(\ref{e2}) with respect to 
$(u_{\mu}, v_{\mu})$; a specific combination of the first order parametric variations turns 
out to be equivalent to a diffusion dynamics of  $U_{\mu}$ along with a drift component:

\begin{eqnarray}
- \; \gamma \left[ 2 
x_{\mu}{\partial \rho\over\partial x_{\mu}} +  
  b_{\mu} {\partial \rho\over\partial b_{\mu}}\right]   =
  {\partial \over \partial U_{\mu}}
\left[ {g_{\mu}\over 2} {\partial \over \partial U_{\mu}} +
 \gamma U_{\mu} \right] \rho
\label{c4}
\end{eqnarray}
where $x_{\mu} \equiv 1- (2- \delta_{\mu}) \; v_{\mu}$, 
$g_{\mu} \equiv g_{kl} = 1+\delta_{kl}$ with $\delta_{kl}=1$ for $k=l$ and $0$ 
for $k \not=l$.

Multiplication of  both sides of eq.(\ref{c4}) with factor 
$\delta(E-\sum U_{\mu})$ and subsequent 
integration over $U$-space gives, along with eq.(\ref{e6+2}), 

\begin{eqnarray}
\gamma \sum_{\mu=1}^{M_0} {\partial P\over\partial z_{\mu}}    =
{\partial \over \partial E} \left[  {\partial \over \partial E} +
\gamma E  \right] P
\label{c5}
\end{eqnarray}
with $z_{\mu} = - {1 \over 2}{\rm ln} \left(|x_{\mu}| \; |b_{\mu} |^2 \right)$,  
$M_0$ as the number of non-zero parameters $x_{\mu}, b_{\mu}$ 
and $\gamma$ as an arbitrary constant with units of $E^{-1}$.

As eq.(\ref{c5}) indicates, the combined effect of first order
parametric variations is  a diffusion of $P(E)$ in the energy space. 
Due to linearity, these first order changes are additive in nature. The 
collective response of the sequence to these changes can then be mimicked by 
the response to a single parameter $Y$:
\begin{eqnarray}
{\partial P \over \partial Y} = 
{\partial  \over \partial E} \left[{\partial  \over \partial E} +  
\gamma E \right] P 
\label{e7}
\end{eqnarray}    
where $Y$ is defined by the condition 
${\partial P \over \partial Y} 
=\sum_{\mu=1}^{M_0} {\partial P\over\partial z_{\mu}}$ 
or, alternatively, 
\begin{eqnarray}
\sum_{\mu=1}^{M_0} {\partial Y\over\partial z_{\mu}} = 1 
\end{eqnarray}
The above condition can easily be solved to give 
\begin{eqnarray}
Y = {1 \over \gamma M_0} \sum_{\mu=1}^M a_{\mu} \; z_{\mu} +c_0
\label{e7+}
\end{eqnarray}
with $M_0 = \sum_{\mu} a_{\mu}$ and $c_0$ is a constant determined by 
the initial condition. Here $a_{\mu}$ are arbitrary constants which can be 
fixed by physical considerations as follows. Eq.(\ref{e7+}) describes $Y$ as a 
a weighted average of $z_{\mu}$s, each representing local accuracy 
fluctuations.  
Assuming no particular bias of folding to any specific error, all $a_{\mu}$s can 
be chosen equal. This gives

\begin{eqnarray}
Y =  -{1 \over 2 \gamma  M_0} \prod'_{\mu} {\rm ln} \left[ |x_{\mu} | 
\; |b_{\mu}|^2 \right] + c_0
\label{e8}
\end{eqnarray}
here $\prod'$ implies a product over non-zero $b_{\mu}$ and $x_{\mu}$ and 
$c_0$ is a constant determined by the initial condition (i.e unfolded sequence). 
(A mathematically rigorous derivation of $Y$ can be found in \cite{psco, ps-valla}).  
Being a function of the system conditions governing folding 
e.g. interaction strengths 
as well as local environment, $Y$ can be termed as the folding parameter.   
During folding, therefore, $P$  undergoes a $Y$-governed diffusion due 
to accuracy driven random forces,  along with a finite drift 
caused by external forces e.g. environmental conditions.

Eq.(\ref{e7}) describes the flow of the probability 
$P(E, Y|E_0, Y_0)$ from an arbitrary initial ensemble of the matrices $H_0$ to 
a steady state (occurring in the limit  ${\partial \rho \over \partial Y} \rightarrow 0$); 
the steady state turns out to be a Gaussian free of any system conditions: 
$P(E, Y \rightarrow \infty) \propto {\rm e}^{-\gamma E^2/2}$.
For an arbitrary initial state $P(E_0, Y_0)$, eq.(\ref{e7}) can be solved to 
give
\begin{eqnarray}
P (E, Y | E_0, Y_0) = c \; {\rm exp}[ - a \; (E- \alpha \; E_0)^2]
\label{e11}
\end{eqnarray}
with $a={1 \over 2(1-\alpha^2)}$, $c={1\over \sqrt{1-\alpha^2}}$, 
and $\alpha={\rm e}^{-(Y-Y_0)}$. 
Let $P(E_0,Y_0)$ represents the energy landscape of the denaturated state. 
The probability $P (E,Y-Y_0)$ for various intermediate stages between 
denaturated and native state can then be obtained by 
integrating  eq.(\ref{e11}) over  $P(E_0,Y_0)$:

\begin{eqnarray}
P(E;Y-Y_0)= \int P(E,Y | E_0, Y_0) P(E_0, Y_0) {\rm d}E_0
\label{e12}
\end{eqnarray}
As eq.(\ref{e8}) indicates, $Y$  increases as  folding proceeds; this is due to increasing contributions from non-zero $u, v$ parameters.   

$P (E,Y-Y_0)$, given by eq.(\ref{e12}),  describes the energy landscape for 
a specific folding stage represented by the functional 
$Y(u, v)$ which contains information about the 
system conditions prevailing during that stage. Thus, beginning from an unfolded sequence, 
the folding process is  governed by the collective 
influence (described by $Y$) of the local interactions (among residues as well as environment) 
on the protein dynamics. Different 
alternatives for pairwise interactions may result in different $Y$ functions and 
therefore many trajectories originating from a given unfolded state. The 
thermodynamic conditions however restrict the choice of the folding trajectories. 
As discussed in the next section, $Y$ dependence of $P(E, Y-Y_0)$ leads to $Y$-governed 
evolution of the thermodynamic measures e.g Gibbs free energy $G$ during folding. The 
thermodynamic stability criterion restricts the native state to occur along the trajectory with 
a well-defined global minimum of $G$ occurring, say at $Y=Y_F$. Due to its dependence  on 
the value of $Y$ and not on its functional form, $G(Y)$ may take a same value at more than one 
trajectory.  Thus folding occurs along trajectories with an approximately similar $G(Y)$ behavior, 
leading to a common global minimum, say at $Y=Y_F$. Existence of a local minimum for   
$Y < Y_F$ may inhibit the folding to a native state. Similarly a local minimum occurring for  $Y > Y_F$ 
may lead to misfolding with changing environmental conditions.

As eq.(\ref{e12}) shows, 
different energy landscapes of initial sequences may lead to different native states. 
However if mutations of some of the residues leave $P(E_0,Y_0)$ of an unfolded 
sequence unchanged, the native state then will also remain unaffected; this is in 
agreement with observed robustness of the native state to sequence-mutations.

The initial ensemble, that is, the ensemble of unfolded or 
fully denatured protein is a linear sequence of residues with 
no secondary of tertiary structure, often existing as a random coil where all 
conformations have comparable energies. $P(E_0,Y_0)$ in this case    
can be described  by a Gaussian distribution: 
\begin{eqnarray}
P(E_0, Y_0) = {1 \over \sqrt{2 \pi \eta }} {\rm e}^{-  {(E_0-\epsilon)^2 \over 2 \eta}}
\label{coil} 
\end{eqnarray}

As clear from their functional forms,  eq.(\ref{c4}) is  also valid for
$\rho_H(H)$ (eq.(\ref{e6})), after replacing
$v_{\mu} \rightarrow \nu_{\mu}, b_{\mu} \rightarrow u_{\mu},
\sum_{k,l, \alpha} \rightarrow \sum_{k, l}$ and
$\rho(U) \rightarrow \rho_H(H)$. Consequently, eq.(\ref{e7}) describes the
the evolution of $P(E)$, given by eq.(\ref{e6+}), too, with corresponding 
changes in $Y$.

\section{Partition Function and Gibbs Free energy $G$}

Eq.(\ref{e12}) for $P(E;Y-Y_0)$ can now be used to obtain the 
partition function $Z$ for the conformation ensemble described by the 
complexity parameter $Y$:  

\begin{eqnarray}
Z(Y-Y_0, T) &=& \int {\rm e}^{-\beta E} P(E, Y-Y_0) \; {\rm d}E  \\
&=& {\sqrt {\pi \over 2 a^2 }} \; {\rm e}^{\beta^2 \over 4 a} \; Z_0(Y_0, \tau) 
\label{e13}
\end{eqnarray}
where $\tau \equiv T/ \alpha$, $\beta=1/kT$ and 
\begin{eqnarray}
Z_0 (Y_0, \tau) = \int {\rm e}^{-\alpha \beta E_0} P(E_0, Y_0) {\rm d}E_0
\label{e13+}
\end{eqnarray}
The free energy of the conformation at 
temperature $T$ is then given by
\begin{eqnarray}
G(Y-Y_0,T) &=& - kT \; {\rm ln} Z   \nonumber \\
&=&  \alpha \; G_0( \tau ) -  {1-\alpha^2 \over 2 k T} - 
k T \; {\rm ln}( \sqrt{2\pi} )  
\label{e14}
\end{eqnarray}
where $G_0(\tau)$ is the free energy of the unfolded state at rescaled 
temperature $\tau $: 
\begin{eqnarray}
G_0(\tau) &=& - k \tau \; {\rm ln} Z_0 
 \label{e14+}
\end{eqnarray}
As eq.(\ref{e14}) implies, the evolution of $G$ at a given temperature $T$ 
is dictated by $\alpha$, and therefore,  $Y$, a function of system conditions through $\{u_{\mu}\}$ and 
$\{v_{\mu}\}$. 
The influence of system conditions on $G$ can then be studied through $Y$.

 The stability of a conformation increases as its $G$ decreases relative to that of 
the unfolded protein.  The thermodynamic stability criterion for folded conformation 
requires its free energy to be minimum. This can be achieved by  
seeking system conditions i.e $\alpha=\alpha_f$ at fixed $T$ for which 
\begin{eqnarray}
{\partial G \over \partial \alpha}|_{{\alpha_f}, T} =0, \qquad \qquad 
{\partial^2 G \over \partial \alpha^2}|_{{\alpha_f}, T} < 0
\label{e15-}
\end{eqnarray}
or equivalently, 
\begin{eqnarray}
\beta \alpha_f + G_0( \beta \alpha_f )+ 
\alpha {\partial G_0(\beta \alpha_f) \over \partial \alpha_f}|_{T}   & =& 0 .  
\label{e15}
\end{eqnarray}
and
\begin{eqnarray}
\alpha_f^2  {\partial G_0^2 \over \partial \alpha_f^2}|_{T}
 - \beta  \alpha_f - 2 G_0( \beta \alpha_f )  < 0
\label{e15-1}
\end{eqnarray}

Substitution of $G_0$ in eq.(\ref{e15}) leads to
complexity parameter  for the thermodynamically stable conformation
at a fixed temperature $T$: $Y_f=Y_0-{\rm ln}(\alpha_f)$.

For example, for an unfolded sequence given by eq.(\ref{coil}),  
eq.(\ref{e13+}) and eq.(\ref{e14+}) give 
\begin{eqnarray}
Z_0(\tau)= {\rm e}^{-(2 \epsilon k \tau - \eta ) \over 2 k^2 \tau^2}, \qquad 
G_0(\tau)= \epsilon - {\eta \over 2 k \tau}.  
\end{eqnarray}
This on substitution in eqs.(\ref{e15}, \ref{e15-1}) gives
\begin{eqnarray}
\alpha_f = {\rm e}^{-(Y_f-Y_0)} = {\epsilon \over (\eta-1) \beta}, \qquad  
{\partial G^2 \over \partial \alpha_f^2}|_{T} =(1-\eta) \beta .  
\end{eqnarray}
The native state can therefore occur only if $\epsilon > 0, \eta >1$.

As eq.(\ref{e14}) implies, the stability of a given conformation changes with 
temperature too. The temperature $T_m$ for maximum stability of a given 
conformation, with all other system conditions fixed, can be obtained by 
the condition ${\partial G \over \partial T}|_\alpha = 0$ which gives
\begin{eqnarray}
2 {\partial G_0(\tau) \over \partial \tau}|_{\alpha} 
+ k \beta^2 (1-\alpha^2) - 2 k \; {\rm log}(\sqrt{2 \pi})  &=& 0   
\end{eqnarray}
or, alternatively, with $S_0(\tau) = -{\partial G_0(\tau) \over \partial \tau}|_{\alpha}$ 
(entropy at a fixed $\alpha$), 
\begin{eqnarray}
2 S_0(\tau) -  k \beta^2 (1-\alpha^2  )
+ 2 k \; {\rm log}( \sqrt{2\pi})   &=& 0 
\label{e15+}
\end{eqnarray}
 By substituting $S_0$ in the above equation, 
one can determine $T_m$ for a specific $\alpha$ i.e a sequence in a specific 
solvent. As eq.(\ref{e15+}) suggests, the stability of a structure decreases 
if the temperature $T > T_m $ or $T < T_m$; this also agrees with the simulation 
studies. For example, for $P(E_0, Y_0)$ given by eq.(\ref{coil}),  
$S_0(\tau) |_{\alpha}=-{\eta  \over 2 k \tau^2}$. Eq.(\ref{e15+}) then 
gives $T_m={(1-\alpha^2) + \eta \alpha^2 \over k^2 {\rm ln}(2 \pi)}$. 
It is easy to check that ${\partial^2 G \over \partial T^2}|_\alpha < 0$ 
at $T=T_m$, indicating a decreasing $G(T)$ and therefore stability 
for $T > T_m$ or $T < T_m $.

Note $\alpha$ (through $Y$ ) depends on both, interactions within sequence as well as with 
the environment (through set $\{u, v \}$). Following our approach, the 
folding  therefore occurs when the matrix $v$ (or $C$ for case eq.(\ref{e6+})),  for a  
specific interaction matrix $u$,  will satisfy eq.(\ref{e15}) with 
$G_0$ of the unfolded state at a temperature $T/\alpha$. The approach also explains the existence of specific  folding pathways at a fixed temperature: $Y_F$ in the parametric 
space $\{u_{\mu}, v_{\mu} \}$ is connected to $Y_0$ through several paths 
however folding occurs along paths with relatively maximum stability (among all paths) 
for an intermediate state too. These folding paths correspond to minimum free energy change between any two intermediate points. 
Also note that $Y$ (through $\{u,v \}$) is evolving with time the rapidity of which depends on the environment; 
the information may help in the determination of folding speed 
at a fixed temperature. 

The appearance of $G_0$ in eqs.(\ref{e14}-\ref{e15+})  indicates that 
the information specifying the native structure as well as the pathway 
to attain that state is contained in the amino 
acid sequence of each protein. The presence of 
both $\alpha$ and $T$ in these equations however reveals the dependence of folding process on environmental 
factors as well as various interactions among residues. Thus  nearly identical amino acid sequences 
may not fold similarly if their environment is different. 
This is in agreement with the results obtained by simulation studies of proteins.

\section{Heat capacity and entropy for denaturation}

Heat capacity $C_p$ (at constant pressure p), defined as   
\begin{eqnarray}
C_p = {\partial {\langle E \rangle} \over \partial T} \vert _p 
= k \beta^2 {\partial^2  ( \beta G) \over \partial \beta^2}\vert_p ,
\label{e16}
\end{eqnarray}
is  an important measure to study the dynamics 
of unfolding \cite{priv} and the hydrophobic effect on the protein stability. 

As eq.(\ref{e15}) and eq.(\ref{e15+}) indicate, the existence of a thermally stable native 
state depends on the specific relation of $G_0$, $Y$ and $T$. It may not 
be satisfied by a sequence under certain environmental conditions; the protein then  
will not fold into its biochemically functional form. 
Further a folded conformation may unfold or "denature" if changes in  system conditions e.g  
temperatures, concentrations of solutes, pH conditions, mechanical forces, and the chemical denaturants result in violation of the eq.(\ref{e15}) or eq.(\ref{e15+}) . The effect of all these changes on $C_p$ can be studied through its $Y$-formulation (obtained from eq.(\ref{e16}) 
and eq.(\ref{e14})):

 \begin{eqnarray}
C_p (\tau) &=&  {k \beta^2 \over \alpha} 
\left[ \beta^2 \alpha^2  {\partial^2  G_0 \over \partial \beta^2} \vert_p + 
2 {\partial  G_0 \over \partial \beta} \vert_p + \alpha (\alpha^2-1) \right]  \\
&=& C_{p0}(\tau) + k \beta^2 (\alpha^2-1)
\label{e17}
\end{eqnarray}
with $C_{p0}(\tau) = k \beta_0^2 {\partial^2  ( \beta_0 G) \over \partial \beta_0^2}\vert_p$ 
as the heat capacity of unfolded protein at temperature $\tau =T/\alpha$ and $\beta_0=1/k \tau$.

The unfolding primarily occurs due to exposure of side chains (e.g. non-polar groups),  
buried in the native state, to solvent. The folding is 
believed to be dominated by the polar groups binding helped by solvent. 
Both these process involve $C_p$ change; for a sequence going from a 
state "$Y_i$" to "$Y_f$"  at temperature $T$, the change in specific heat  
$\Delta C_p \equiv C_p(\tau_f)-C_p(\tau_i)$ can be given as 
(from eq.(\ref{e17})), 
\begin{eqnarray}
\Delta C_p &=&  C_{p0}(\tau_f) -C_{p0}(\tau_i) + 
k \; \beta^2 \; (\alpha^2_f - \alpha^2_i)
\label{e18}
\end{eqnarray}
with $\alpha_k = {\rm e}^{-(Y_k-Y_0)}$ for $k=f, i$.   
Due to positive and negative $C_p$ of hydration for apolar and polar groups, respectively, the 
sign of $\Delta C_p$ can provide information about nature of solvation e.g polar or apolar, 
and folding/ unfolding/ misfolding etc. For example, for the unfolded state given by 
eq.(\ref{coil}), 
\begin{eqnarray}
\Delta C_p &=&  k \; \beta^2 \; (1-\eta) (\alpha^2_f - \alpha^2_i)
\label{e19}
\end{eqnarray}
Thus for unfolding, which corresponds to $Y_f < Y_i$ or $\alpha_f > \alpha_i$,  one gets $\Delta C_p >0$. 
The folding, with $Y_f > Y_i$, similarly corresponds to $\Delta C_p < 0$.

 The entropy 
\begin{eqnarray}
S &=&  k \; \left( {\rm ln} Z + \beta E \right) 
= k \; \beta^2 \; {\partial  G \over \partial \beta}  
\label{e20}
\end{eqnarray} 
is another important 
thermodynamic property commonly measured for proteins. A competition of entropy 
with stabilizing forces determines the possibility of unfolding which occurs 
at temperatures when $S$ becomes dominant. The $Y$-dependence of $S$ can be given as
\begin{eqnarray}
S(Y,T) &=&  S_0 (\tau) - (1/2) k \; \beta^2 \; (1-\alpha^2) + k \; {\rm ln}\sqrt{ 2\pi}  
\label{e21}
\end{eqnarray} 
with $S_0(\tau)$ as the entropy of unfolded sequence at temperature $\tau$.

The entropy change $\Delta S$ contains information about 
reversibility ($\Delta S \le 0$) 
or irreversibility ($\Delta S > 0$) of the process. $Y$- dependence 
of $\Delta S$ can then be used to determine the system-condition which can lead to refolding 
of a misfolded protein. For example, presence of chaperon molecules may change interaction 
parameters (referred by $u$)  and therefore $Y$ and $\Delta S$.  
For a sequence changing from state $Y_i \rightarrow Y_f$,  $\Delta S $ is  
\begin{eqnarray}
\Delta S = S_f-S_i &=&  S_0(\tau_f) -S_0(\tau_i) + 
(1/2) k \beta^2 (\alpha^2_f - \alpha^2_i). 
\label{e22}
\end{eqnarray}  
For unfolded state given by eq.(\ref{coil}), we get   

\begin{eqnarray}
\Delta S =k \beta^2 (1+\eta)(\alpha_f^2 -\alpha_i^2)/2
\label{e23}
\end{eqnarray}  
which implies an 
increase of entropy for $Y_f < Y_i$ (unfolding) and a decrease of entropy 
for $Y_f > Y_i$ (folding).

The simulation studies suggest that the ratio of the entropy change, $\Delta S $, 
to the heat capacity change, $\Delta C_p$, for the dissolution of a variety of 
hydrophobic compounds is a constant. This is confirmed by our formulation 
too. The ratio can be determined from 
eq.(\ref{e16}) and eq.(\ref{e20}):
 
\begin{eqnarray}
{\Delta S \over \Delta C_p} &=& {\partial^2 \over \partial \beta^2} 
{\rm ln} {Z_f \over Z_i } 
\label{e24}
\end{eqnarray}
It is easy to see, from eq.(\ref{e19}) and eq.(\ref{e23}), that 
the ratio depends only on the properties of unfolded sequence: 
${\Delta S \over \Delta C_p} =  {1\over 2} {1+\eta \over 1-\eta} $.

\section{conclusion}

To summarize, a protein sequence in general is described by a multi-parametric ensemble of interactions. Our study shows however that the thermodynamic properties of the sequence 
are governed by a single parameter (besides temperature) which is basically a measure of average 
uncertainty associated with the local interactions.  
 The formulation  
provides an analytical understanding of some important observations  obtained by 
computer simulation studies of proteins e.g dependence of native state on 
original sequence, the role of solvent, decrease of stability of the native state 
above and below the critical temperature.   The stability of folded sequence 
against mutations can also be explained by the $Y$-dependence of free energy: 
a mutation may change the interaction parameter $u_{kl}$ however $Y_F$ may 
remain unaffected (change being averaged out in the combination of interaction 
parameters). Such mutations will leave native state unaffected. 
The $Y$-formulation also  explains the selection of specific folding pathways among 
infinite number of possibilities and can be used to identify them. 
We have yet to apply it to  many other simulation studies observations, 
for example, the observed preference to the functionality and folding speed , instead of stability,  
as the main criteria for selection of a natural protein conformation, studies on 
misfolding of proteins etc.

	The random matrix approach described here is applicable only for
the cases of interaction matrix with independent matrix elements; this takes 
into account only two-body interactions. In general, a protein is a complex 
system with many body interactions and consequently the interaction matrix 
contains correlated elements. The generalization of single parametric 
formulation to protein models  with correlated matrix elements is very desirable; 
we intend to pursue some of these 
questions in future studies.

\end{document}